\documentclass{article}

\usepackage{arxiv}

\usepackage[utf8]{inputenc} 
\usepackage[T1]{fontenc}    
\usepackage{lmodern}        
\usepackage{hyperref}       
\usepackage{url}            
\usepackage{booktabs}       
\usepackage{amsfonts}       
\usepackage{nicefrac}       
\usepackage{microtype}      
\usepackage{graphicx}
\usepackage{natbib}

\title{A Basket Trial Design Based on Power Priors}

\author{
    Lukas Baumann
   \\
    Institute of Medical Biometry \\
    University of Heidelberg \\
  69120 Heidelberg, Germany \\
  \texttt{baumann@imbi.uni-heidelberg.de} \\
   \And
    Lukas Sauer
   \\
    Institute of Medical Biometry \\
    University of Heidelberg \\
  69120 Heidelberg, Germany \\
  \texttt{} \\
   \And
    Meinhard Kieser
   \\
    Institute of Medical Biometry \\
    University of Heidelberg \\
  69120 Heidelberg, Germany \\
  \texttt{} \\
  }

\usepackage{amsmath}
\usepackage{bbm}
\usepackage{bm}
\usepackage{enumerate}
\usepackage{algorithm}
\usepackage{algpseudocode}
\usepackage{pifont}
\usepackage{makecell}
\usepackage{booktabs}
\usepackage{longtable}
\begin{document}
\maketitle

\begin{abstract}
In basket trials a treatment is investigated in several subgroups. They are primarily used in oncology in early clinical phases as single-arm trials with a binary endpoint. For their analysis primarily Bayesian methods have been suggested, as they allow partial sharing of information based on the observed similarity between subgroups. \citet{fujikawa2020} suggested an approach using empirical Bayes methods that allows flexible sharing based on easily interpretable weights derived from the Jensen-Shannon divergence between the subgroup-wise posterior distributions. We show that this design is closely related to the method of power priors and investigate several modifications of Fujikawa's design using methods from the power prior literature. While in Fujikawa's design, the amount of information that is shared between two baskets is only determined by their pairwise similarity, we also discuss extensions where the outcomes of all baskets are considered in the computation of the sharing weights. The results of our comparison study show that the power prior design has comparable performance to fully Bayesian designs in a range of different scenarios. At the same time, the power prior design is
computationally cheap and even allows analytical computation of operating characteristics in some settings.
\end{abstract}

\LTcapwidth=\textwidth

\section{Introduction}
\label{sec:introduction}

Basket trials are used when a treatment is investigated in several subgroups which are called baskets. They are primarily used in oncology, where the baskets comprise patients with different tumor histologies but all patients in the trial share a common molecular marker or a genetic
mutation \citep{hirakawa2018}. Most basket trials are early-phase single-arm trials that investigate a
binary endpoint such as tumor response. While many basket trials are simply analyzed by either pooling the data of all baskets or analyzing each basket independently \citep{hobbs2022}, these approaches have obvious limitations. When data are pooled across baskets, it is not possible to distinguish active from inactive baskets. On the other hand, analyzing the baskets independently
leads to low power.

To overcome these shortcomings, a range of different approaches for analyzing basket trials has been suggested in recent years. The key idea of these designs is that information is shared between baskets to increase the power \citep{pohl2021}. For that purpose, most designs use Bayesian tools which enable partial sharing of information depending on observed similarity between baskets. Among these tools are Bayesian hierarchical modeling \citep{berry2013, neuenschwander2016} and model averaging \citep{psioda2021, hobbs2018}

A different idea, that can be considered an empircal Bayes approach, was proposed by \citet{fujikawa2020}. In their design, baskets are at first individually modeled using beta-binomial models. The posterior distributions including the shared information are beta
distributions with posterior parameters that are calculated as weighted sums of the individual posterior parameters, where the weights are derived from a divergence measure that quantifies the pairwise similarity between baskets. The approach is attractive, as the amount of information that is shared is very flexible and transparent through the weights. The design is also computationally cheap compared to most other Bayesian basket trial designs, as the weights are easy to compute and thus posterior distributions can be calculated analytically even for a large number of baskets. In fact, it is possible to analytically calculate the operating characteristics such as type 1 error rate (TOER) and power, at least for a moderate number of baskets and with at most one interim analysis.

While Fujikawa et al. suggest to use the Jensen-Shannon divergence (JSD) between the individual posterior distributions to derive the weights that determine the amount of information that is shared, an obvious question is whether the performance of the design can be improved by using different strategies to find the weights. To address this question, we establish a connection between Fujikawa's basket trial design and the approach of power priors, which was originally proposed
for borrowing information from historical data \citep{ibrahim2000}. We show that many ideas from the power prior literature can also be applied to basket trials and thus to extend Fujikawa's design. Furthermore, we compare these design  to other basket trial designs in a comprehensive investigation including both analytical calculations and simulations. Comparing different basket trial designs is challenging, due to the high number of parameters and design elements that can be adapted
(e.g. number and sample sizes of baskets, type and number of interim assessments, prior and tuning parameters of the different designs) and the huge number of different scenarios that could be considered. Another issue is control of the TOER. The posterior probability threshold for declaring efficacy can be calibrated such that the family wise error rate (FWER) is protected at a certain level under the global null hypothesis. However, this may still lead to high TOER inflation under mixed scenarios where some baskets are active and some are inactive. Therefore, simply comparing the power of different designs may not be the best approach. Finally, choosing the comparator designs is difficult. A substantial number of different basket trials designs have been suggested in the last years (see \citealp{pohl2021} for an overview). However, there are hardly any comparisons between them
available and thus it is currently unclear which designs perform best in which cases.

The rest of this article is structured as follows: In Section \ref{sec:methods}, the setting of the basket trials is described and the designs which are included in our investigation are presented. Section \ref{sec:compstudy} outlines the framework of the comparison study, the results are shown in Section \ref{sec:results}. We conclude with a discussion of the findings.

\section{Methods}
\label{sec:methods}

\subsection{Setup and Notation}

The basic setting is an uncontrolled single-stage phase II basket trial, where a binary endpoint is investigated in $K \geq 2$ disjoint subgroups, the baskets, with sample sizes $n_1, ..., n_K$. Let $\bm{p} = (p_1, ..., p_K)$ be the vector of response rates, $\bm{R} = (R_1, ..., R_K)$ the random vector representing the number of responses in each basket and $\bm{r} = (r_1, ..., r_K)$ the vector of realisations of $\bm{R}$.
The treatment under investigation is considered efficacious in a basket $k \in \{1, ..., K\}$, if $p_k$ is greater than $p_0$, a prespecified null response probability.

\citet{pohl2021} identify four components of basket trial designs: Information sharing, interim futility assessment, interim efficacy assessment and the final analysis, whereby the two interim assessment components are optional. The objective of this article is to compare the information sharing components of different designs. Therefore, although different futility and efficacy assessment strategies are often proposed as part of basket trial designs, in the rest of this section we will only discuss the information sharing part of the designs and skip the two optional interim assessment components. For the final analysis, different strategies exist. Since all designs considered in this article use Bayesian methodology, decisions are made based on the posterior
distribution of $p_k$ given the observed data. We use the final decision rules proposed in the respective designs, which is that, given a prespecified probability threshold $\lambda \in (0, 1)$, basket $k$ is declared as active if $\mathbb{P}(p_k > p_0| \bm{R} = \bm{r}) \geq \lambda$ for Fujikawa's design and $\mathbb{P}(p_k > p_0| \bm{R} = \bm{r}) > \lambda$ for all
other designs. Since our newly proposed sharing method is built on Fujikawa's design, we also use Fujikawa's decision rule.

\subsection{Fujikawa's Design}

In the design of \citet{fujikawa2020}, the baskets are at first analyzed individually using a beta-binomial model, i.e. given a prior distribution $\text{Beta}(s_{1,k}, s_{2,k})$ for $p_k$, the
posterior distribution for basket $k \in \{1, ..., K\}$ is:
\begin{equation*}
	\pi(p_k|R_k = r_k) = \text{Beta}(s_{1,k} + r_k, s_{2,k} + n_k - r_k)
\end{equation*}
The posterior distribution with information sharing is then:
\begin{equation*}
	\pi(p_k|\bm{R} = \bm{r}) = \text{Beta}(
	\sum_{i = 1}^K \omega_{k,i} (s_{1,k} + r_i), 
	\sum_{i = 1}^K \omega_{k,i} (s_{2,k} + n_i - r_i))
\end{equation*}
\citet{fujikawa2020} suggest to estimate the weights $\omega_{k,i}$ based on a function of the
pairwise JSD between the posterior distributions of the individual baskets without sharing any information. Specifically, the weights are computed as:
\begin{equation*}
	\omega_{k,i} = \begin{cases}
		(1 - \text{JSD}(\pi(p_k|R_k = r_k), \pi(p_i|R_i = r_i))^\varepsilon & 
		\text{if } (1 - \text{JSD}(\pi(p_k|R_k = r_k), \pi(p_i|R_i = r_i))^\varepsilon > \tau \\
		0 & \text{otherwise}
	\end{cases},
\end{equation*}
where $\text{JSD}(P_k, P_i)$ is the JSD between two distributions $P_k$ and $P_i$, which is given by \citep{fuglede2004}: 
\begin{equation*}
	\text{JSD}(P_k, P_i) = \frac{1}{2}\text{KLD}\left(P_k, \frac{1}{2}(P_k + P_i)\right) +
	\frac{1}{2}\text{KLD}\left(P_i, \frac{1}{2}(P_k + P_i)\right).
\end{equation*}
KLD refers to the Kullback-Leibler divergence, which is given by:
\begin{equation*}
	\text{KLD}(P_k, P_i) = \int f_k(x) \log \frac{f_k(x)}{f_i(x)} dx,
\end{equation*}
where $f_k$ and $f_i$ are the densities of $P_k$ and $P_i$, respectively. Through the KLD, the formula for the JSD involves a logarithm. Fujikawa et al. use the natural logarithm, which gives an upper bound of $\ln(2)$ for the JSD. However, we use the base 2 logarithm instead, since this results in weights that are bounded between 0 and 1 (\citealp{lin1991}; see also \citealp{baumann2022}).

\begin{figure}
	\centering
	\includegraphics[width=0.8\textwidth]{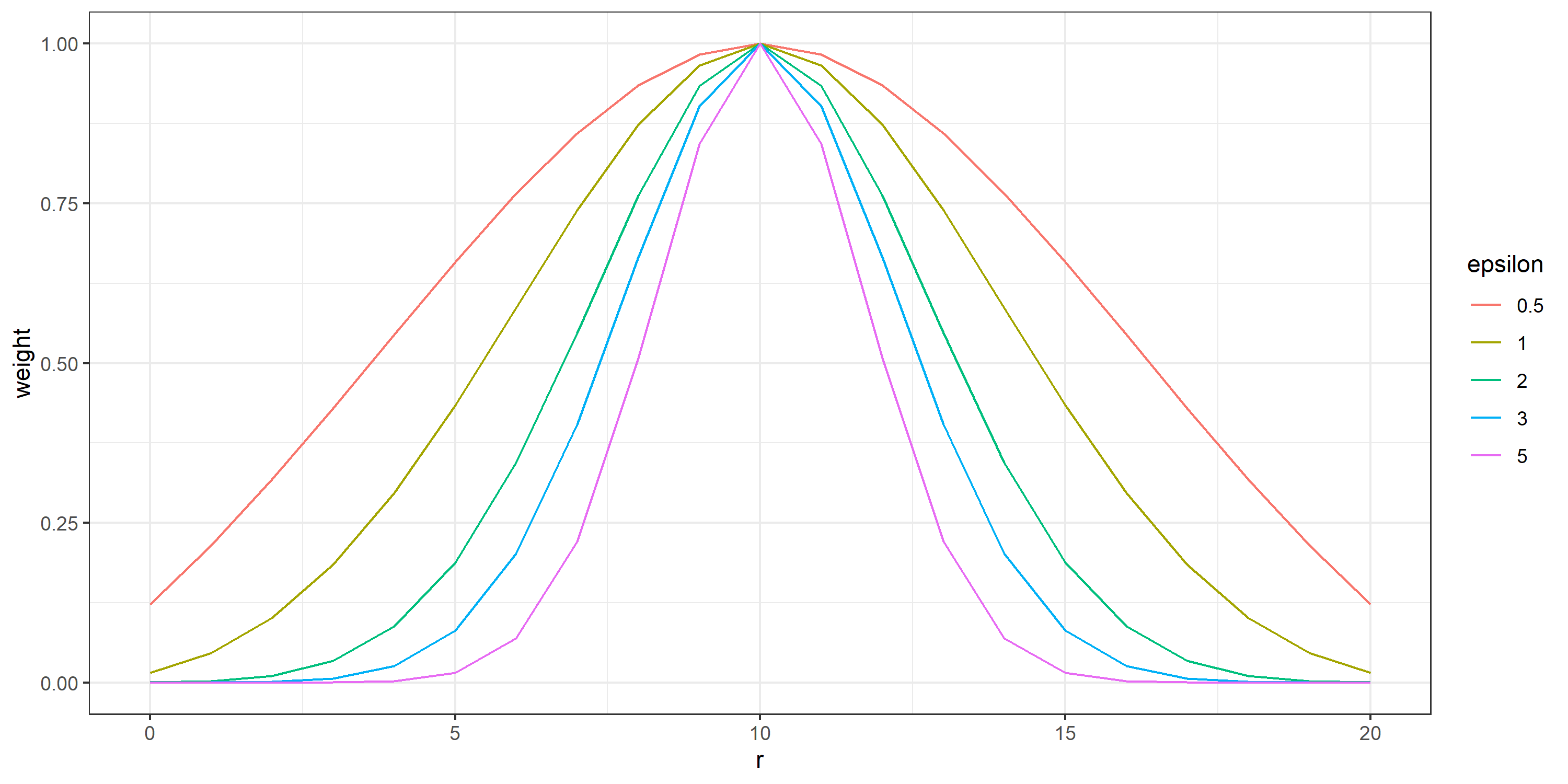}
	\caption{Different JSD weights between two baskets with $n = 20$, when the number of responses in one basket is $r_1 = 10$ with varying values for the tuning parameter $\varepsilon$}
	\label{fig:fujikawaweights}
\end{figure}

The JSD weights depend on two tuning parameters $\varepsilon > 0$ and $\tau \in [0, 1]$. $\varepsilon$ determines how fast the weights decline when the posterior distributions become more different. If $\varepsilon$ is small, then a lot of information is shared even when the individual posterior distributions are different. As $\varepsilon$ increases, weights decrease when the results of two baskets are less similar. Figure \ref{fig:fujikawaweights} shows how the weights change for different choices of $\varepsilon$. $\tau$ establishes a minimum of similarity that is necessary for any information to be shared between two baskets. Note that the JSD is 0 if and only if the two distributions are identical \citep{briet2009}. Hence, (given equal prior parameters) in the case of unequal sample sizes weights are always smaller than 1 even if the response rates are identical.

Fujikawa's design is computationally very cheap, such that operating characteristics can even be computed analytically for $K \leq 5$ baskets and at most one interim analysis in reasonable time using the R-package baskexact. The computationally most expensive part in the evaluation of the posterior distribution is calculating the weights. However, since Fujikawa's design shares information based only on pairwise similarity, the number of possible weights that can occur is
limited. In the simplest case of a single-stage design with equal sample sizes, there are only at most $(n + 1)^2$ different possible weights.

\subsection{Power Priors}

Power priors were initially proposed by \citet{ibrahim2000} as a tool to incorporate historical data in a study. Based on an initial prior $\pi_0(\theta)$, the historical data $D_0$, its likelihood $L(\theta|D_0)$ and a power parameter $\omega \in [0,1]$, the power prior for a parameter $\theta$ is defined as:
\begin{equation*}
	\pi(\theta|D_0,\omega) \propto L(\theta|D_0)^\omega \pi_0(\theta).
\end{equation*}
The power parameter can either be treated as a random variable with its own prior distribution or as a fixed value. If the value is fixed, then for binary data and an initial $\text{Beta}(s_1, s_2)$ prior for the response probability $p$ of the current study, the posterior distribution has the following closed form:
\begin{equation*}
	\pi(p|R = r,R_0 = r_0,\omega) = \text{Beta}(s_1 + r + \omega \cdot r_0, s_2 + (n - r) + \omega(n_0 - r_0)),
\end{equation*}
where $R$, $r$ and $n$ are the random variable of responses, its realization and the sample size of the current study, and $R_0$, $r_0$ and $n_0$ are the respective variable and quantities of the
historical study. Hence, the historical data are weighted by the power parameter $\omega$. The power prior can also be used if multiple historical studies are available. Let $\bm{R_0} = (R_{0,1}, ..., R_{0,H})$ be the random vector of responses of $H$ historical data sets with realizations
$\bm{r_0} = (r_{0,1}, ..., r_{0,H})$ and sample sizes $\bm{n_0} = (n_{0,1} ..., n_{0,H})$. Let further
$\bm{\omega} = (\omega_1, ..., \omega_H)$ be the vector of weights. Then the extended power prior is:
\begin{equation*}
	\pi(\theta|\bm{R_0},\bm{\omega}) \propto \left(\prod_{i=1}^H L(\theta|D_{0,i})^{\omega_i}\right) \pi_0(\theta)
\end{equation*}
For binary data, this leads to the following posterior distribution:
\begin{equation*}
	\pi(p|\bm{R_0} = \bm{r_0}, \bm{\omega}) = \text{Beta}(
	s_1 + \sum_{i = 1}^H \omega_i \cdot r_{0,i}, 
	s_2 + \sum_{i = 1}^H \omega_i (n_{0,i} - r_{0,i})).
\end{equation*}
When for a basket $k$ the data of all other baskets $j \in \{1,...,K\}\backslash k$ are treated as the historical data in the original power prior approach, a slight change in notation results in an application of the power prior to basket trials:
\begin{equation*}
	\pi(p_k|\bm{R} = \bm{r}, \bm{\omega}) = \text{Beta}(
	s_{1,k} + \sum_{i = 1}^K \omega_{k,i} r_i, 
	s_{2,k} + \sum_{i = 1}^K \omega_{k,i} (n_i - r_i)).
\end{equation*}
with $\omega_{k, k} = 1$. The similarity to Fujikawa's design is obvious. In Fujikawa's design data are shared in the same way. The only difference is that in Fujikawa's design information from the prior distribution is also shared between baskets, since the prior parameters are also included in the weighted sum. We would argue that sharing prior information may be desirable when the prior is derived from observed data of earlier studies. However, if non-informative priors are used, then we think that the power prior formulation is favorable. Of course, if a prior with very small $s_1$ and $s_2$ is used, e.g. $\text{Beta}(0.01, 0.01)$, then the difference between the two approaches will be negligible.

We will now discuss several different approaches to derive the weights. When we describe the weights, we will always use notation that relates to the application in basket trials, even when methods derived from the literature on borrowing data from historical studies are discussed.

\subsection{Calibrated Power Prior Weights}

A flexible way to derive the weights, that was suggested in the context of borrowing from historical data, was proposed by \citet{yuan2017} which they termed the calibrated power prior (CPP). Their idea is to calculate the weights based on the Kolmogorov-Smirnov (KS) statistic between the current and
the historical study. Note that for two binary data sets, the KS statistic is equal to the absolute rate difference. The KS statistic $S_{KS; k,i}$ is scaled by defining $S_{k,i} = \max(n_k, n_i)^{1/4} S_{KS;k,i}$. The weights are then given by:
\begin{equation*}
	\omega_{k,i} = \frac{1}{1 + \exp{(a + b \log{(S_{k,i})})}},
\end{equation*}
where $a \in \mathbb{R}$ and $b > 0$ are tuning parameters. $b > 0$ is required to have weights that are strictly monotonically decreasing in the rate difference.

\begin{figure}
	\centering
	\includegraphics[width=0.99\textwidth]{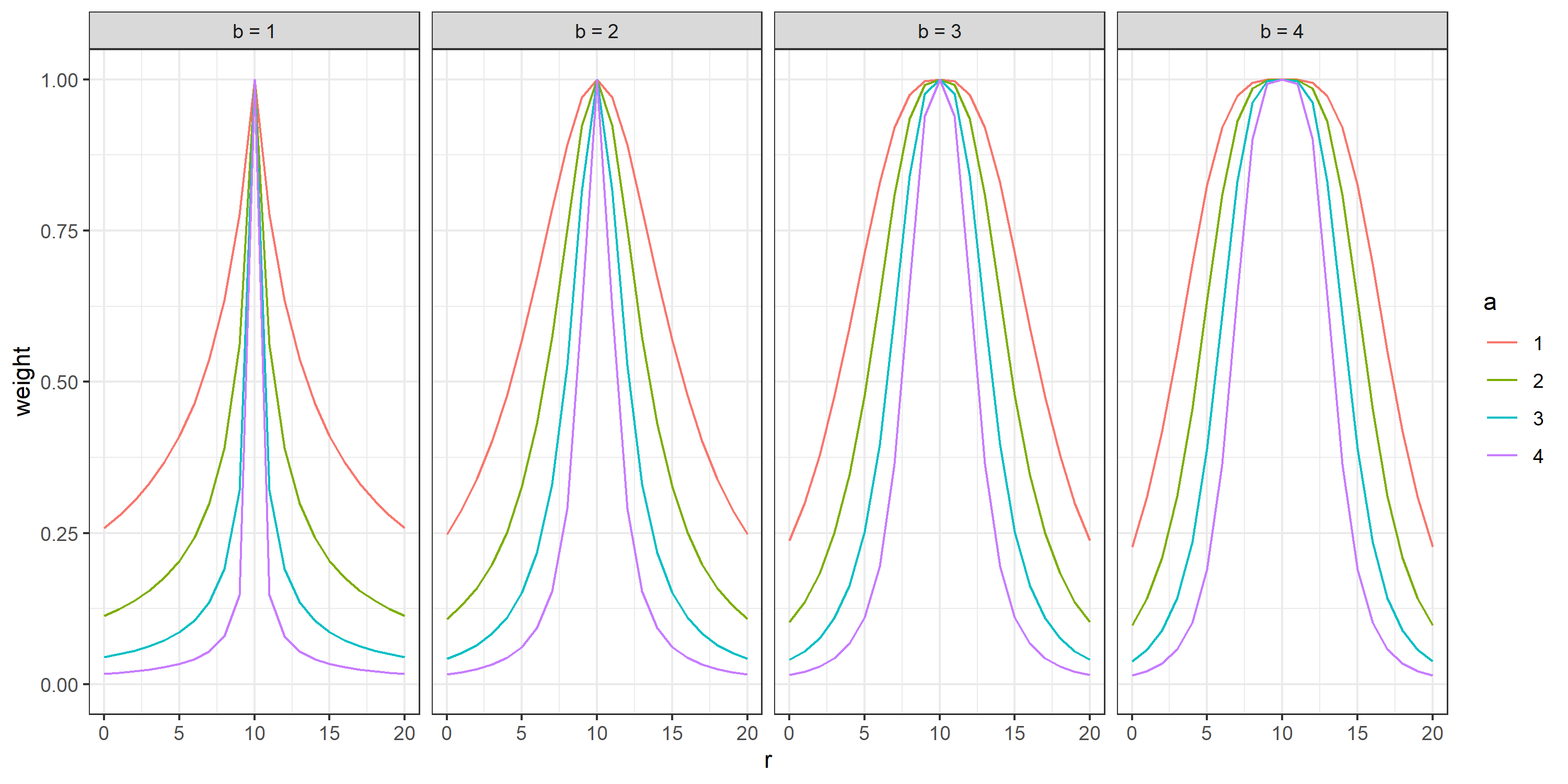}
	\caption{Different CPP weights between two baskets with $n = 20$, when the number of responses in one basket is $r_1 = 10$ with varying values for the tuning parameters $a$ and $b$}
	\label{fig:cppweights}
\end{figure}

While \citet{yuan2017} only consider the case of a single historical study, their approach can
easily be extended to basket trials by computing the weights based on all pairwise comparisons between baskets, equivalently to computing the pairwise weights based on the JSD in Fujikawa's design. The shape of the weight functions is a little more flexible than in Fujikawa's design,
since the CPP weights include two tuning parameters that determine the shape of the weight function, see Figure \ref{fig:cppweights}. Since, other than the JSD weights, the value of the KS statistic only depends on the rates of the two binary samples, the CPP weights are also equal
to 1 if the response rates are equal but the sample sizes differ.

Note that \citet{yuan2017} also give an algorithm on how to select the two tuning parameters based
on the available historical data. However, since for basket trials no data is available before conducting the trial, we suggest to select $a$ and $b$ based on simulations under different scenarios.

\subsection{Maximum Marginal Likelihood Weights}

While Fujikawa's design and the application of the CPP to basket trials both calculate the weights based on the pairwise similarity between baskets only, \citet{gravestock2019} find that in the context
of borrowing from multiple historical studies, using weights that are only based on pairwise similarity between the current and all historical studies is not the best approach. Instead, they suggest to choose the vector of weights as the value that maximizes its marginal likelihood,
with the constraint $\bm{\omega} \in [0, 1]^K$. This has the advantage that the weights are based on the information of all historical studies at once. This can also be adapted to basket trials but is
computationally more expensive since in a trial with $K$ baskets the marginal likelihood has to be maximized $K$ times, once for each basket as the "current study" and all other baskets as the
"historical studies". The marginal likelihood of the weights $\bm{\omega_k}$ for a certain basket $k$ is:
\begin{equation*}
	L(\bm{\omega_k}|\bm{r}) = {n_k \choose r_k}
	\frac{B\left(r_k + s_{1,k} + \sum_{i \in \mathcal{K}} \omega_{k,i} r_i, n_k - r_k + s_{2,k} + \sum_{i \in \mathcal{K}} \omega_{k,i} (n_i - r_i)\right)}
	{B\left(s_{1,k} + \sum_{i \in \mathcal{K}} \omega_{k,i} r_i, s_{2,k} + \sum_{i \in \mathcal{K}} \omega_{k,i} (n_i - r_i)\right)},
\end{equation*}
where $B$ is the beta function and $\mathcal{K} = \{1, ... K\}\backslash k$. This is the density of a beta-binomial distribution. \citet{gravestock2019} note that the constraint $\bm{\omega_k} \in [0, 1]^K$ is necessary and that, for binomial data, there is no closed form for the maximum of the marginal likelihood and thus the maximum has to be found using numerical optimization methods.

\subsection{Weights Based on Global Heterogeneity}

An extension of pairwise weights based on the JSD or the CPP approach that considers the ideas of \citet{gravestock2019} to incorporate the entire available information, and not just pairwise similarity, to calculate the weights, is as follows: Compute global weights $\omega^\star \in [0, 1]$ derived from the heterogeneity of the results of all baskets and then use $\omega^\prime_{k,i} = \omega_{k,i} \cdot \omega^\star$ for determining the amount of shared information. For example, as an
extension of Fujikawa's design, $\omega^\star$ can be calculated based on the generalized JSD for $K > 2$ distributions as:
\begin{equation*}
	\omega^\star = \left(1 - \text{JSD}(\pi(p_1|R_1 = r_1), ..., \pi(p_K|R_K = r_K)\right)^{\varepsilon^\star},
\end{equation*}
where $\varepsilon^\star$ is a tuning parameter. The generalized JSD is (in the case of equally weighted distributions) defined as:
\begin{equation*}
	\text{JSD}(\bm{P}) = \frac{1}{K} \sum_{i = 1}^K \text{KLD}(P_i, \bar{P}),
\end{equation*}
where $\bm{P}$ is a vector of distributions and $\bar{P} = \frac{1}{k}\sum_{i=1}^K P_i$. Note that when the logarithms to compute the KLD are base $K$, then $0 \leq \text{JSD}(\bm{P}) \leq 1$ \citep{nielsen2021}. The amount of information that is shared between two baskets would then depend on the pairwise similarity through the pairwise JSD, and also on the overall heterogeneity through the generalized JSD of all baskets. Therefore, if overall heterogeneity is high, less information is shared between two baskets, even if they have identical posterior distributions.

Since numerical integration is necessary to compute the JSD and since the number of possible weights is much higher than when just calculating pairwise weights, using these global weights slows down computation significantly and makes analytical computation of the operating characteristics infeasible. We therefore suggest another global heterogeneity function $h$ based on the following considerations. Assume for simplicity that $K = 3$ and let $rr_i = r_i / n$ be the response rates. Further define $d_1 = rr_{(3)} - rr_{(2)}$ and $d_2 = rr_{(2)} - rr_{(1)}$ as the differences of the ordered response rates. Our function $h(d_1, d_2)$ should then satisfy:
\begin{enumerate}
	\item $h(0, 0) = 1$, i.e. no heterogeneity is defined as observing equal response rates
	\item $h(0.5, 0.5) = 0$, i.e. maximum heterogeneity is defined as observing response rates that are an equidistant sequence from 0 to 1
\end{enumerate}
The following function satisfies these two conditons:
\begin{equation*}
	h(\bm{d}) = (1 - \sum_{i=1}^{K-1} d_i \cdot 10^{-\sum_i (d_i - 1/(K - 1))^2})^{\varepsilon^\star},
\end{equation*}
where $\bm{d} = (d_1, ..., d_K)$ and $\varepsilon^\star$ is a tuning parameter. This function is much cheaper to compute than the JSD and also allows analytical computation of operating characteristics.

$\omega^\star$ could also be set to a fixed value smaller than 1 to limit the maximum amount of borrowing between two baskets to give more weight to the data actually observed in the respective basket. This is similar to the idea of the EXNEX design described in Section \ref{sec:exnex}.

\subsection{Bayesian Hierarchical Model Design}

In the Bayesian Hierarchical Model (BHM) design of \citet{berry2013}, the response rates $p_k$ are logit-transformed and assumed to follow a common normal distribution, i.e.
\begin{equation*}
	\theta_k = \log{\left(\frac{p_k}{1-p_k}\right)} - \log{\left(\frac{p_{\text{targ},k}}{1 - p_{\text{targ},k}}\right)}
\end{equation*}
and
\begin{equation*}
	\theta_k \sim N(\mu, \sigma^2).
\end{equation*}
The target values $p_{\text{targ},k}$ are added such that baskets with different target response rates can be modeled together. To complete the model, prior distributions for $\mu$ and $\sigma^2$ must be specified. The prior distribution for $\mu$ is a normal distribution with an expected value that corresponds to the null hypothesis and a large variance. For $\sigma^2$ \citet{berry2013} use an inverse-gamma prior, but other distributions such as half-normal, half-t or uniform can also be used \citep{cunanan2019, neuenschwander2016}.

\subsection{EXNEX Design}
\label{sec:exnex}

As an extension of the simple BHM, \citet{neuenschwander2016} proposed the EXNEX (exchangeabilty-nonexchangebility) design, which models the logit-transformed response rates $\theta_k = \log{\left(p_k / (1-p_k)\right)}$ as a mixture of a BHM (exchangeability part) and individual distributions for each basket (nonexchangeability part). A fixed mixture weight $w$ (potentially different for different baskets) has to be specified to complete the model. Neuenschwander et al. note that even though $w$ is fixed, the posterior distribution is a mixture with updated weights (see e.g. \citealp{bolstad2016}).

\subsection{Bayesian Model Averaging Design}

In the Bayesian Model Averaging (BMA) design by \citet{psioda2021}, $J$ models $M_1, ..., M_J$, representing all possible cluster structures, are evaluated and then weighted by their posterior probability. For example, in the case of $K = 3$ there are five different models. All response rates can either be equal ($p_1 = p_2 = p_3$) or different ($p_1 \neq p_2 \neq p_3$), resulting in a single cluster or three clusters, respectively. The other possibilities are $p_1 = p_2 \neq p_3$, $p_1 = p_3 \neq p_2$ and $p_1 \neq p_2 = p_3$, each resulting in two clusters. In each cluster, the data are pooled and modeled using a beta-binomial model. Let $D_j$ be the number of distinct response rates in model $M_j$, then the prior for $M_j$ is\footnote{Note that in Psioda et al.'s paper there is a typo in this formula (as confirmed by the first author). However, the results in their paper are based on the formula given here, which is also implemented in the R package bmabasket.}:
\begin{equation*}
	\pi(M_j) \propto \exp(D_j \cdot \psi),
\end{equation*}
where $\psi$ is a tuning parameter. \citet{psioda2021} only explore the use of values $\psi \geq 0$, but values less than 0 can also be used. If $\psi = 0$ then each model has the same prior probability,
$\psi > 0$ gives higher prior weights to models with more clusters, and with $\psi < 0$ higher prior probabilities are assigned to models with fewer clusters.

The posterior probabilities for $M_j$ have a closed form solution, so the posterior probabilities for $\bm{p}$ also have an analytical solution. However, as $K$ increases the number of different models and therefore computational time also increase.

\section{Comparison Study}
\label{sec:compstudy}

In the previous section we presented different options to derive weights for a basket trial design based on a power prior design or Fujikawa's design. There are three general approaches (weights based on pairwise similarity only, weights based on pairwise and global similarity, weights based on the maximum of the marginal likelihood) that can be modified and combined in various ways. We conducted a comparison study where we investigate the following designs:

\begin{itemize}
	\item BHM design
	\item BMA design
	\item Power prior design with pairwise weights based on the CPP approach
	\item Power prior design with pairwise weights based on the CPP approach and global weights based on the heterogeneity function $h$ (termed CPP-Global)
	\item Power Prior  design with pairwise weights based on the CPP approach and a fixed global weight (termed CPP-Nex)
	\item EXNEX design
	\item Fujikawa's design
	\item Power prior design with pairwise weights and global weights based on the JSD (termed JSD-Global)
\end{itemize}

The BHM, BMA and EXNEX design were chosen as competing designs as these are among the few basket trial designs that are implemented in an R package available on CRAN. The BHM and EXNEX design
are implemented in bhmbasket \citep{bhmbasket2022} and the BMA design is implemented in bmabasket \citep{bmabasket2022}.

\subsection{Setup}

As already discussed, we consider a single-stage trial to focus on the information sharing aspect of basket trials. As a further simplification, we assume equal sample sizes of $n = 20$ in all baskets. The number of baskets is set to $K = 4$. Guided by the scenarios investigated by \citet{berry2013} and \citet{broglio2022} we considered 7 different scenarios presented in Table \ref{tab:scenarios}.

\begin{table}[]
	\centering
	\caption{Scenarios considered in the comparison study}
	\label{tab:scenarios}
	\begin{tabular}{@{}lllll@{}}
		\toprule
		Scenario & Basket 1 & Basket 2 & Basket 3 & Basket 4 \\
		\midrule
		Global Null           & 0.15 & 0.15 & 0.15 & 0.15 \\
		Global Alternative    & 0.4  & 0.4  & 0.4  & 0.4  \\
		One in the Middle     & 0.4  & 0.4  & 0.3  & 0.5  \\
		Linear                & 0.15 & 0.25 & 0.35 & 0.45 \\
		Good Nugget           & 0.15 & 0.15 & 0.15 & 0.4  \\
		Bad Nugget            & 0.15 & 0.4  & 0.4  & 0.4 \\
		Half                  & 0.15 & 0.15 & 0.4  & 0.4  \\
		\bottomrule
	\end{tabular}
\end{table}

To find the optimal tuning parameter values for each design, we followed the steps of a simulation study by \citet{broglio2022}. At first, a set of potential prior and tuning parameter values is selected for each method. Then, for each method and each combination of different parameter values, the probability threshold $\lambda$ is chosen such that the one-sided FWER $\alpha$ is protected at 5\% under the global null hypothesis. For each combination of parameter values and the threshold $\lambda$ found in the previous step, the expected number of correct decisions (ECD) is calculated under all the scenarios in Table \ref{tab:scenarios}. The tuning parameter values resulting in the
highest mean ECD over all scenarios are then chosen as optimal and the results under the different scenarios of each method using the optimal tuning parameter values are then further investigated by computing rejection rates and FWERs. Posterior means were also computed and are presented in the Supplementary Material.

For Fujikawa's design and the power prior design with CPP, CPP-Global, and CPP-Nex weights, all operating characteristics are calculated analytically using the R-Package baskexact (\url{https://github.com/lbau7/baskexact}). For all other methods, results are based on the same 10,000 simulated datasets per scenario. BHM and EXNEX additionally require MCMC sampling. 10,000 samples plus 1000 burn-in samples were used. For this, the R-package basksim (\url{https://github.com/lbau7/basksim}) was used, which implements the power prior design with weights that are not implemented in baskexact. Additionally, basksim provides wrappers for functions of the bmabasket and bhmbasket package to have a unified syntax to compare the operating
characteristics. Note that a minimally modified version of bhmbasket was used for the simulations (\url{https://github.com/lbau7/bhmbasket}).

\subsection{Potential Tuning Parameter Values}

For defining the set of potential tuning parameter values we also mostly follow \citet{broglio2022}. For each method, some of the parameters that do not guide the amount of shared information and are considered to have little impact on the performance were held fixed and for the relevant prior and tuning parameters, a grid of potential values was defined.

For Fujikawa's design and all power prior weights, the beta priors were set to $\text{Beta}(1, 1)$. For Fujikawa's design, values for $\tau$ between 0.1 and 0.5 in steps of 0.1 were used, for $\varepsilon$ values between 0.5 and 3 in steps of 0.5, resulting in 30 different combinations of tuning parameters. For the generalized Fujikawa's design, additionally values for $\varepsilon$, also between 0.5 and 3 in steps of 0.5 were considered, leading to 180 different value combinations. In the designs with CPP weights, again, values between 0.5 and and 3 in steps of 0.5 were considered for the two tuning parameters $a$ and $b$, resulting in 36 different combinations. For CPP-Global, the same range of values was used for $\varepsilon^\star$, so in total 216 values were investigated. For
CPP-Nex, fixed values between 0.1 and 0.9 in steps of 0.1 were used for $\omega^\star$, i.e. 288 different combinations were investigated. No tuning parameters are involved in the computation of the MML weights.

For the BHM design, the prior distribution for $\mu$ is fixed at $N(-1.3291, 100)$. The mean of this distribution corresponds to the null hypothesis, given a null response probability of 0.15 and a common target response probability of 0.4. For the prior for $\sigma^2$, a half-normal distribution was used, as this is the prior implemented in bhmbasket. 8 different values equidistant between 0.125 and 2 for the scale parameter of the half-normal distribution, denoted $\phi$, were examined.

For the EXNEX design, additional to the expected value $\mu$ of the common normal distribution, we have $K$ prior distributions for the individual response rates $\theta_k$. For all of these, the normal prior distribution was fixed at $N(-1.7346, 100)$. Note that the mean is different from the BHM mean, since EXNEX does not adjust for the target response rate in the definition of the $\theta_k$. For $\sigma^2$ in the exchangeability part the same distribution with the same values as for the BHM were analyzed. For the additional weight parameter $w$, values 0.1 and 0.9 in steps of 0.1 were considered, which leads to a total of 72 prior parameter value combinations.

For the BMA design, the beta priors were also fixed at $\text{Beta}(1, 1)$. For the only other prior parameter $\psi$, we investigated values between -4 and 4 in steps of 0.5.

\section{Results}
\label{sec:results}

Table \ref{tab:optparam} shows the best prior and tuning parameter values in terms of the mean ECD for all methods. Additional results showing the influence of the choice of parameter values are shown in the Supplementary Material.

\begin{table}[]
	\centering
	\caption{Optimal prior and tuning parameter values for all methods}
	\label{tab:optparam}
	\begin{tabular}{@{}lll@{}}
		\toprule
		Method   & Parameter                     & Value \\ \midrule
		BHM      & $\phi$                        & 0.661 \\
		BMA      & $\psi$                        & -2    \\
		CPP      & $a$                           & 2     \\
		& $b$                           & 1.5   \\
		CPP-Global  & $a$                           & 1.5   \\
		& $b$                           & 1     \\
		& $\varepsilon$                 & 0.5   \\
		CPP-Nex  & $a$                           & 2     \\
		& $b$                           & 2     \\
		& $\omega^\star$                & 0.8   \\
		EXNEX    & $\phi$                        & 0.393 \\
		& $w$                           & 0.6   \\
		Fujikawa & $\varepsilon$                 & 1.5   \\
		& $\tau$                        & 0     \\
		JSD-Global  & $\varepsilon$                 & 0.5   \\
		& $\varepsilon^\star$           & 3     \\
		& $\tau$                        & 0     \\ \bottomrule
	\end{tabular}
\end{table}

Table \ref{tab:optres} shows the results of all methods using the best tuning parameter values in terms of the ECDs. While CPP-Nex has the highest mean ECD by a very small margin, all methods have very similar mean ECD and only differ in the second decimal place. Looking at the individual scenarios, the highest ECDs are naturally achieved in scenarios where all baskets are either active or inactive and borrowing works best. The Linear scenario has the lowest ECDs for all methods,
with values around 3.

BHM has average performance across all scenarios, with a mean ECD almost identical to EXNEX. EXNEX has the highest ECDs in the Global Alternative, One in the Middle and the Linear scenario, but the
CPP-based methods have very similar values in these scenarios. In the Good Nugget, the Bad Nugget and the Half scenarios, the power prior design with MML weights performed best, with a large advance in the Half scenario compared to all other methods. However, on the other hand, the MML method is quite far behind in the Global Alternative and the One in the Middle scenario.

Fujikawa's design and the other power prior methods, except for MML, perform relatively similar. The CPP weights slightly improved the mean ECD compared to Fujikawa's design. Additionally adding the global weights based on the proposed heterogeneity function $h$ has almost no effect on the CPP weights. Adding global JSD weights slightly decreases the mean ECD as compared to Fujikawa's design. An improvement is only seen in the Good Nugget scenario.

\begin{table}[]
	\centering
	\caption{ECDs under all scenarios using the optimal tuning parameter values for each method. The best result per column is bold.}
	\label{tab:optres}
	\begin{tabular}{@{}lllllllll@{}}
		\toprule
		Method & \makecell{Global \\ Null} & \makecell{Global \\ Alt}    & \makecell{One in the \\ Middle} & Linear & \makecell{Good \\ Nugget} & \makecell{Bad \\ Nugget} & Half & Mean           \\ 
		\midrule
		BHM      & 3.928          & 3.865 & 3.734 & 2.999 & 3.442          & 3.468          & 3.365          & 3.543          \\
		BMA      & 3.904          & 3.871 & 3.719 & 2.964 & 3.342          & 3.451          & 3.319          & 3.510          \\
		CPP      & 3.916          & 3.910 & 3.817 & 3.066 & 3.403          & 3.497          & 3.321          & 3.561          \\
		CPP-Global  & 3.922          & 3.909 & 3.819 & 3.056 & 3.410          & 3.486          & 3.323          & 3.561          \\
		CPP-Nex  & 3.919          & 3.910 & 3.816 & 3.066 & 3.420          & 3.494          & 3.336          & \textbf{3.566} \\
		EXNEX    & 3.917          & \textbf{3.922}& \textbf{3.834}    & \textbf{3.083} & 3.343       & 3.515      & 3.243            & 3.551 \\
		Fujikawa & 3.908          & 3.882 & 3.738 & 3.068 & 3.340          & 3.520          & 3.352          & 3.544          \\
		JSD-Global  & 3.925          & 3.878 & 3.731 & 2.906 & 3.476          & 3.414          & 3.325          & 3.522          \\
		MML      & \textbf{3.932} & 3.640 & 3.469 & 2.985 & \textbf{3.489} & \textbf{3.528} & \textbf{3.527} & 3.510          \\ \bottomrule
	\end{tabular}
\end{table}

Looking at the rejection rates and FWERs in Table \ref{tab:rejrates}, in the Global Alternative scenario the results look quite similar for all methods except for MML, which shares less information between the baskets. In the One in the Middle Scenario, relevant differences in power can be seen for example in the third basket, were for MML the power is 67\% and ranges from 82\% to 88\% for the other methods. A similar variance in power is seen in the second basket with $p = 0.25$ of the Linear scenario and in the active basket in the Good Nugget scenario. However, TOERs also have to be considered.

In the scenarios with at least one inactive basket the FWERs are far above the nominal level for most methods. In the Bad Nugget scenario, the TOER for the inactive basket increases to more than 30\% for some methods. Adding the global weights in the JSD-Global method led to relevant differences compared to Fujikawa's design in the Good Nugget and the Half scenarios. In the Good Nugget scenario JSD-Global has a FWER that is lower by 5 percentage points while at the same time increasing the power by 5 percentage points compared to Fujikawa's design. In the Half scenario the FWER is 6 percentage points lower in JSD-Global, but also the power is lower by 5 percentage points. The three CPP variations have almost identical rejection rates in all scenarios. Hence, the increased complexicity does not lead to the desired improvement. BHM also increases the FWERs considerably, but less so than EXNEX and the other methods mentioned above. Note that in the Global Null scenario BHM slightly exceeds the significance level of \(5\%\) due to simulation error resulting from the MCMC sampling. BMA also tends to have less type 1 error inflation, but is also among the methods with the lowest power, especially in the Good Nugget scenario.

As already seen in Table \ref{tab:optres}, MML stand out, as it has significantly less type 1 error inflation. MML has the highest FWER in the Good Nugget scenario at 16\%. In the Bad Nugget scenario, were FWERs are especially high for the other methods with values between 27\% and 34\%, MML has a FWER of only 12\%. On the other hand, of course, this goes hand in hand with a loss in power. In the Global Alternative scenario, MML has basketwise power values that are more than 5 percentage points below the other methods. The highest difference is seen in the One in the Middle scenario, where for the third basket the power is only at 67\% with the MML method and at 88\% for the CPP methods. In the Good Nugget and the Half scenario, however, the power of MML is compareable to the other methods, while showing far lower FWER than the other methods in the Half scenario.

\begin{longtable}[]{ccccccc}
	\caption{Rejection rates under all scenarios using the optimal tuning parameter values for each method}
	\label{tab:rejrates} \\
	\toprule
	Scenario & Method & Basket 1 & Basket 2 & Basket 3 & Basket 4 & FWER  \\ \midrule
	Global Null 
	& BHM      & 0.020    & 0.018    & 0.020    & 0.018    & 0.052 \\
	& BMA      & 0.024    & 0.023    & 0.024    & 0.024    & 0.049 \\
	& CPP      & 0.021    & 0.021    & 0.021    & 0.021    & 0.048 \\
	& CPP-Global  & 0.019    & 0.019    & 0.019    & 0.019    & 0.048 \\
	& CPP-Nex  & 0.020    & 0.020    & 0.020    & 0.020    & 0.049 \\
	& EXNEX    & 0.022    & 0.020    & 0.021    & 0.020    & 0.049 \\
	& Fujikawa & 0.023    & 0.023    & 0.023    & 0.023    & 0.048 \\
	& JSD-Global  & 0.019    & 0.018    & 0.020    & 0.019    & 0.049 \\
	& MML      & 0.018    & 0.016    & 0.017    & 0.017    & 0.049 \\
	\midrule
	Global Alt 
	& BHM      & 0.965    & 0.969    & 0.966    & 0.968   & . \\
	& BMA      & 0.967    & 0.970    & 0.968    & 0.967   & . \\  
	& CPP      & 0.977    & 0.977    & 0.977    & 0.977   & . \\
	& CPP-Global  & 0.977    & 0.977    & 0.977    & 0.977   & . \\
	& CPP-Nex  & 0.978    & 0.978    & 0.978    & 0.978   & . \\
	& EXNEX    & 0.980    & 0.982    & 0.979    & 0.980   & . \\
	& Fujikawa & 0.970    & 0.970    & 0.970    & 0.970   & . \\
	& JSD-Global  & 0.970    & 0.972    & 0.968    & 0.968   & . \\
	& MML      & 0.907    & 0.912    & 0.910    & 0.910   & . \\
	\midrule
	One in the Middle 
	& BHM      & 0.960    & 0.958    & 0.826    & 0.995   & . \\
	& BMA      & 0.955    & 0.955    & 0.812    & 0.996   & . \\
	& CPP      & 0.972    & 0.972    & 0.877    & 0.996   & . \\
	& CPP-Global  & 0.972    & 0.972    & 0.878    & 0.996   & . \\
	& CPP-Nex  & 0.971    & 0.971    & 0.877    & 0.996   & . \\
	& EXNEX    & 0.979    & 0.978    & 0.879    & 0.998   & . \\
	& Fujikawa & 0.959    & 0.959    & 0.824    & 0.996   & . \\
	& JSD-Global  & 0.965    & 0.953    & 0.818    & 0.995   & . \\
	& MML      & 0.906    & 0.904    & 0.673    & 0.980   & . \\ 
	\midrule
	Linear 
	& BHM      & 0.194    & 0.483    & 0.781    & 0.928    & 0.194 \\
	& BMA      & 0.240    & 0.492    & 0.781    & 0.931    & 0.240 \\
	& CPP      & 0.247    & 0.566    & 0.805    & 0.942    & 0.247 \\
	& CPP-Global  & 0.245    & 0.558    & 0.805    & 0.939    & 0.245 \\
	& CPP-Nex  & 0.248    & 0.564    & 0.808    & 0.942    & 0.248 \\
	& EXNEX    & 0.284    & 0.597    & 0.828    & 0.938    & 0.284 \\
	& Fujikawa & 0.236    & 0.553    & 0.807    & 0.944    & 0.236 \\
	& JSD-Global  & 0.245    & 0.462    & 0.762    & 0.927    & 0.245 \\
	& MML      & 0.092    & 0.391    & 0.760    & 0.926    & 0.092 \\
	\midrule
	Good Nugget
	& BHM      & 0.060    & 0.063    & 0.066    & 0.628    & 0.144 \\
	& BMA      & 0.076    & 0.077    & 0.080    & 0.575    & 0.152 \\
	& CPP      & 0.075    & 0.075    & 0.075    & 0.629    & 0.154 \\
	& CPP-Global  & 0.072    & 0.072    & 0.072    & 0.627    & 0.152 \\
	& CPP-Nex  & 0.077    & 0.077    & 0.077    & 0.651    & 0.161 \\
	& EXNEX    & 0.085    & 0.086    & 0.089    & 0.607    & 0.174 \\
	& Fujikawa & 0.087    & 0.087    & 0.087    & 0.602    & 0.178 \\
	& JSD-Global  & 0.065    & 0.057    & 0.060    & 0.658    & 0.129 \\
	& MML      & 0.059    & 0.060    & 0.061    & 0.669    & 0.159 \\
	\midrule
	Bad Nugget 
	& BHM      & 0.272    & 0.910    & 0.915    & 0.915    & 0.272 \\
	& BMA      & 0.269    & 0.904    & 0.907    & 0.908    & 0.269 \\
	& CPP      & 0.322    & 0.940    & 0.940    & 0.940    & 0.322 \\
	& CPP-Global  & 0.322    & 0.936    & 0.936    & 0.936    & 0.322 \\
	& CPP-Nex  & 0.323    & 0.939    & 0.939    & 0.939    & 0.323 \\
	& EXNEX    & 0.338    & 0.947    & 0.950    & 0.949    & 0.338 \\
	& Fujikawa & 0.288    & 0.936    & 0.936    & 0.936    & 0.288 \\
	& JSD-Global  & 0.302    & 0.899    & 0.908    & 0.910    & 0.302 \\
	& MML      & 0.116    & 0.881    & 0.880    & 0.883    & 0.116 \\
	\midrule
	Half
	& BHM      & 0.139    & 0.134    & 0.821    & 0.817    & 0.224 \\
	& BMA      & 0.158    & 0.157    & 0.818    & 0.816    & 0.222 \\
	& CPP      & 0.179    & 0.179    & 0.839    & 0.839    & 0.278 \\
	& CPP-Global  & 0.173    & 0.173    & 0.835    & 0.835    & 0.270 \\
	& CPP-Nex  & 0.178    & 0.178    & 0.846    & 0.846    & 0.276 \\
	& EXNEX    & 0.220    & 0.221    & 0.841    & 0.837    & 0.332 \\
	& Fujikawa & 0.176    & 0.176    & 0.852    & 0.852    & 0.274 \\
	& JSD-Global  & 0.143    & 0.144    & 0.808    & 0.805    & 0.210 \\
	& MML      & 0.080    & 0.079    & 0.844    & 0.843    & 0.144 \\ \bottomrule
\end{longtable}

Mean posterior means for all methods and all scenarios are presented in
the Supplementary Material.

\section{Discussion}

The basket trial design proposed by \citet{fujikawa2020} offers the possibility to flexibly share information between baskets depending on the observed similarity while still being computationally cheap compared to many other Bayesian basket trial designs. We showed how Fujikawa's design is
related to the Bayesian power prior method and suggested several modifications and extensions, mostly based on other ideas from the power prior and basket trial literature. Specifically, we suggested extensions that incorporate the entire available information and not just the pairwise similarity to compute the weights that determine the amount of information that is shared between baskets. We investigated Fujikawa's design and the power prior design with five different weights and compared them to the BHM, the EXNEX and the BMA design, after selecting the optimal prior and tuning parameter values for all methods based on the highest mean ECD across seven different scenarios in a single-stage trial with four baskets.

Using the optimal tuning parameter values, all methods showed relatively similar performance in terms of the mean ECD. In specific scenarios, however, differences in terms of ECD, TOERs and power were seen. Specifically, the power prior design with MML weights had a much less type 1 error rate of at most 16\% in the Good Nugget scenario, whereas the highest FWERs were seen in the Bad Nugget scenario with values between 27\% and 34\% for all other methods. However, due to the trade-off between power and TOER, the MML also had the lowest power in the scenarios where all baskets are active.

The proposed JSD-Global weights that take the overall heterogeneity into account resulted in slightly lower type 1 error inflation compared to Fujikawa's design in some scenarios with only a negligible decrease in the mean ECD. Incorporation of a new heterogeneity measure $h$, however, had no such effect as compared to the CPP weights that are only based on pairwise similarity. A fixed global weight in combination with the CPP weights had the best mean ECD of all methods numerically, but
there were no notable differences to the pairwise CPP weights. This is in line with the the results of \citet{broglio2022}, who also observed no relevant differences in mean performance between simple and complex models.

As there is a large number of tuning and design parameters in a basket trial to consider, some simplifications were necessary and consequently our comparison study has some limitations for practical applications. We assumed equal sample sizes in all baskets, although variance in sample
sizes between different subgroups can be high in practice. Further research is needed to investigate how different methods perform under different sample size scenarios and how weights should be selected in the power prior design in this case.

Furthermore, we considered a single-stage design with no interim analyses, due to the focus on the different sharing techniques. In real life, basket trials will usually have some kind of interim assessment to stop for futility and/or efficacy to reduce trial time and expected sample sizes. \citet{fujikawa2020} suggested a futility assessment based on the posterior predictive probability, while \citet{berry2013} and \citet{psioda2021} use the posterior probability for interim decisions. Both approaches can be used in the power prior basket design. Comparing different methods in combination with different interim analysis strategies is an interesting and practically relevant topic for further research, but introduces several more parameters that have to be considered in a comparison study.

A further extension would be to consider different constraints than control of the FWER under the global null hypothesis when selecting the tuning parameter values. We saw that FWERs under mixed alternative scenarios were quite diverse, which makes it hard to directly compare different methods in terms of the rejection rates. However, for Bayesian basket trail designs, there are currently no methods to control the FWER in the strong sense (i.e. under all mixed alternative scenarios) and in
an early phase trial this may not even be desirable. As an alternative, \citet{kaizer2021} suggested that a weighted sum of the FWERs under different scenarios could be controlled instead. A different approach is to optimize a utility function instead of looking at TOERs and power separately \citep{jiang2021}.

In conclusion, we have shown that the power prior design is a flexible and computational cheap method that performs similar to other Bayesian basket trial designs under a range of different scenarios. Taking only the mean ECD into account, weights based on the CPP approach by \citet{yuan2017} showed the best performance; if high type 1 error inflation is a concern, then MML weights can be used.

\section*{Funding}

This work was supported in part by the German Research Foundation under grant KI 708/9-1.

\bibliographystyle{apalike}
\bibliography{references}

\clearpage
\newpage

\hypertarget{supplementary-material}{%
\section*{Supplementary Material}\label{supplementary-material}}
\addcontentsline{toc}{section}{Supplementary Material}

\setcounter{section}{0}
\setcounter{page}{1}
\setcounter{table}{0}
\setcounter{figure}{0}

\renewcommand{\thetable}{S\arabic{table}}  
\renewcommand{\thefigure}{S\arabic{figure}}

\section{Sensitivity Analyis on Tuning Parameter}

In our comparison study, tuning parameter values that maximize the mean ECD across seven different scenarios were selected. However, different values may be optimal for a different set of scenarios. To assess the sensitivity of the choice of tuning parameter values, optimal parameters were selected for only a subset of the initially defined scenarios. Namely, for the sensitivity analysis the One in the Middle and the Linear scenario were ignored, as often only a single alternative response rate is considered. Results in terms of the ECDs are presented in Table \ref{tab:optres_sens}.

\begin{table}[H]
\centering
\caption{ECDs under scenarios with a common alternative response probability using the optimal tuning parameter values for each method. The best result per column is bold.}
\label{tab:optres_sens}
\begin{tabular}{@{}lllllll@{}}
\toprule
& \makecell{Global \\ Null} & \makecell{Global \\ Alt}    & \makecell{Good \\ Nugget} & \makecell{Bad \\ Nugget} & Half & Mean           \\ 
\midrule
BHM      & 3.938 & 3.779 & 3.497 & 3.501 & 3.400 & 3.623 \\
BMA      & 3.926 & 3.768 & 3.465 & 3.463 & 3.397 & 3.604 \\
CPP      & 3.919 & 3.779 & 3.482 & 3.527 & 3.433 & 3.628 \\
CPP-Global  & 3.930 & 3.777 & \textbf{3.503} & 3.507 & 3.410 & 3.625 \\
CPP-Nex  & 3.930 & \textbf{3.813} & 3.483 & 3.520 & 3.409 & \textbf{3.631} \\
EXNEX    & \textbf{3.937} & 3.771 & 3.498 & 3.518 & 3.400 & 3.625 \\
Fujikawa & 3.920 & 3.780 & 3.434 & \textbf{3.540} & 3.405 & 3.616 \\
JSD-Global  & 3.930 & 3.803 & 3.502 & 3.425 & 3.377 & 3.607 \\
MML      & 3.932 & 3.640 & 3.489 & 3.528 & \textbf{3.527} & 3.623 \\
\bottomrule
\end{tabular}
\end{table}

Results remain similar in the sense that all methods perform relatively similar and all methods only differ in the second decimal place. Again, CPP-Nex has the highest mean ECD by a very small margin. However, now CPP-Nex wins in the Global Alternative scenario, CPP-Gen has the highest ECD in the Good Nugget scenario and Fujikawa performs best in the Bad Nugget scenario, albeit again only by a very small margin. MML is still superior in the Half scenario. Note that the results for MML did not change as it includes no tuning parameters but are shown again for better comparability.

Table \ref{tab:optparam2} shows how the optimal parameters changed in
comparison to the optimal parameters for the full set of scenarios. For
many methods the optimal tuning parameters for the limited set of
scenarios are very similar, often the tuning parameter values only
changed by one position in the considered grid of values. A bigger
change is seen in the EXNEX design, where a smaller value for the
scale-parameter \(\phi\) was optimal when all scenarios were considered.
A further interesting change is seen in JSD-Global where the optimal
\(\tau\) is 0.3 in the limited set of scenarios, whereas \(\tau = 0\)
was optimal for both JSD-Global and Fujikawa's design in the full set of
scenarios.

\begin{longtable}[]{llll}
\caption{Optimal prior and tuning parameter values for all methods. Values (1) refers to the tuning parameter values that result in the highest mean ECD across all 7 scenarios and Values (2) to the tuning parameter values achieving the highest mean ECD across the selected scenarios with a common alternative response probability.}
\label{tab:optparam2} \\
\toprule
Method   & Parameter                     & Value (1)   & Value (2) \\ \midrule
BHM      & $\phi$                        & 0.661       & 0.929 \\
BMA      & $\psi$                        & -2          & -1 \\
CPP      & $a$                           & 2           & 2.5 \\
         & $b$                           & 1.5         & 1.5 \\
CPP-Global  & $a$                           & 1.5         & 2 \\
         & $b$                           & 1           & 1 \\
         & $\varepsilon$                 & 0.5         & 0.5 \\
CPP-Nex  & $a$                           & 2           & 2.5 \\
         & $b$                           & 2           & 2.5 \\
         & $\omega^\star$                & 0.8         & 0.6 \\
EXNEX    & $\phi$                        & 0.393       & 0.929 \\
         & $w$                           & 0.6         & 0.7 \\
Fujikawa & $\varepsilon$                 & 1.5         & 2 \\
         & $\tau$                        & 0           & 0.1 \\
JSD-Global  & $\varepsilon$                 & 0.5         & 1 \\
         & $\varepsilon^\star$           & 3           & 2.5 \\
         & $\tau$                        & 0           & 0.3 \\ \bottomrule
\end{longtable}

\begin{figure}
\centering
\includegraphics[width=0.65\textwidth]{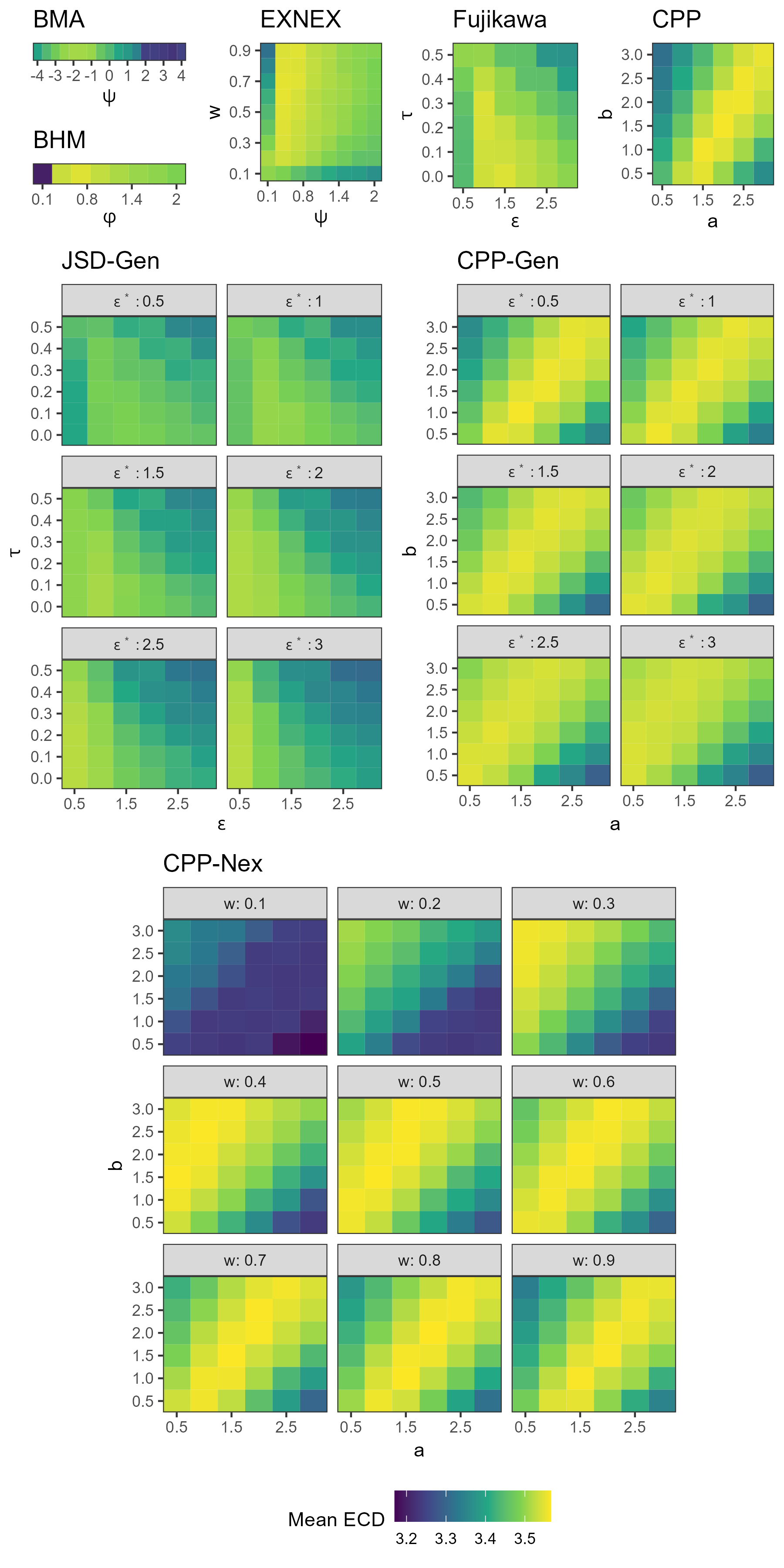}
\caption{Heatmap showing the influence of tuning and prior parameter values on the mean ECD across all seven scenarios.}
\label{fig:heatplot}
\end{figure}

Figure \ref{fig:heatplot} shows how the choice of tuning parameter values affects the mean ECD across all seven scenarios. For most methods there are many different choices of tuning parameter values that result in a high mean ECD.

We further explored how results change when values for the tuning parameter values are not selected by their mean performance across all scenarios, but based on their ECD in one specific scenario. We
considered the Bad Nugget, Half and Linear scenario for that, since these are the most interesting scenarios as there is a trade-off between TOER control and power. The Good Nugget scenario also has an active and three inactive baskets. However, when only one basket is active the designs are trivially optimised by borrowing as little information as possible in the respective design. ECDs are shown in Table \ref{tab:optres3}, parameter values are shown in Table \ref{tab:optparam3}. Improvements in the Linear and the Bad Nugget scenario are only minor. In the Half scenario, ECDs improved a lot compared to the results using tuning parameters chosen based on the mean ECD. Changes in tuning parameter values were also the largest in this scenario, tending towards values that share the least information. However, at the cost of an improvement in the Half scenario the mean ECDs reduced significantly for many methods. Only CPP-Nex remained above 3.5 mean ECD. Note that MML still achieves the highest ECD in the Half scenario (compare with the results in Table \ref{tab:optres_sens}).

\begin{longtable}[]{llllllllll}
\caption{ECDs under all scenarios using the tuning parameter values that are optimal for the Linear, Bad Nugget, and Half scenario.}
\label{tab:optres3} \\
\toprule
\makecell{Optimal \\ Scenario} & Method & \makecell{Global \\ Null} & \makecell{Global \\ Alt}    & \makecell{One in the \\ Middle} & Linear & \makecell{Good \\ Nugget} & \makecell{Bad \\ Nugget} & Half & Mean           \\ 
\midrule
Linear 
& BHM & 3.907 & 3.951 & 3.882 & 3.059 & 3.302 & 3.411 & 3.210 & 3.532 \\ 
& BMA & 3.835 & 3.959 & 3.870 & 3.009 & 3.003 & 3.349 & 3.051 & 3.439 \\ 
& CPP & 3.908 & 3.928 & 3.840 & 3.088 & 3.343 & 3.484 & 3.293 & 3.555 \\ 
& CPP-Global & 3.890 & 3.976 & 3.935 & 3.071 & 3.225 & 3.338 & 3.097 & 3.504 \\ 
& CPP-Nex & 3.891 & 3.979 & 3.942 & 3.078 & 3.208 & 3.322 & 3.046 & 3.495 \\ 
& EXNEX & 3.884 & 3.926 & 3.818 & 3.101 & 3.204 & 3.525 & 2.966 & 3.489 \\ 
& Fujikawa & 3.878 & 3.973 & 3.916 & 3.111 & 3.124 & 3.356 & 2.996 & 3.479 \\ 
& JSD-Global & 3.906 & 3.941 & 3.839 & 2.922 & 3.379 & 3.356 & 3.260 & 3.515 \\ 
\midrule
Bad Nugget 
& BHM & 3.938 & 3.779 & 3.618 & 2.955 & 3.497 & 3.501 & 3.400 & 3.527 \\ 
& BMA & 3.926 & 3.768 & 3.587 & 2.920 & 3.465 & 3.463 & 3.397 & 3.504 \\ 
& CPP & 3.919 & 3.779 & 3.619 & 2.985 & 3.482 & 3.527 & 3.433 & 3.535 \\ 
& CPP-Global & 3.922 & 3.766 & 3.594 & 2.963 & 3.486 & 3.513 & 3.438 & 3.526 \\ 
& CPP-Nex & 3.921 & 3.810 & 3.649 & 3.017 & 3.468 & 3.535 & 3.406 & 3.544 \\ 
& EXNEX & 3.895 & 3.891 & 3.762 & 3.089 & 3.281 & 3.566 & 3.111 & 3.514 \\ 
& Fujikawa & 3.924 & 3.794 & 3.611 & 3.007 & 3.412 & 3.541 & 3.396 & 3.526 \\ 
& JSD-Global & 3.930 & 3.803 & 3.605 & 2.850 & 3.502 & 3.425 & 3.377 & 3.499 \\ 
\midrule
Half
& BHM & 3.941 & 3.706 & 3.513 & 2.866 & 3.540 & 3.438 & 3.415 & 3.488 \\ 
& BMA & 3.948 & 3.001 & 2.858 & 2.638 & 3.598 & 3.244 & 3.429 & 3.245 \\ 
& CPP & 3.935 & 3.388 & 3.183 & 2.733 & 3.556 & 3.300 & 3.451 & 3.364 \\ 
& CPP-Global & 3.937 & 3.431 & 3.209 & 2.735 & 3.563 & 3.309 & 3.448 & 3.376 \\ 
& CPP-Nex & 3.924 & 3.765 & 3.596 & 2.962 & 3.485 & 3.510 & 3.445 & 3.527 \\ 
& EXNEX & 3.941 & 3.687 & 3.491 & 2.872 & 3.546 & 3.453 & 3.417 & 3.487 \\ 
& Fujikawa & 3.931 & 3.600 & 3.366 & 2.895 & 3.484 & 3.490 & 3.435 & 3.457 \\ 
& JSD-Global & 3.935 & 3.610 & 3.378 & 2.749 & 3.493 & 3.396 & 3.449 & 3.430 \\ 
\bottomrule
\end{longtable}

\begin{table}[H]
\centering
\caption{Prior and tuning parameter values for all methods that are optimal under the Linear, Bad Nugget, and Half scenario}
\label{tab:optparam3}
\begin{tabular}{@{}lllll@{}}
\toprule
Method   & Parameter                     & Linear   & Bad Nugget & Half \\ \midrule
BHM      & $\phi$                        & 0.393    & 0.929      & 2    \\
BMA      & $\psi$                        & -3.5     & -1         & 4    \\
CPP      & $a$                           & 2        & 2.5        & 3    \\
         & $b$                           & 2        & 1.5        & 0.5  \\
CPP-Global  & $a$                           & 1.5      & 3          & 2.5  \\
         & $b$                           & 2.5      & 2          & 0.5  \\
         & $\varepsilon$                 & 0.5      & 0.5        & 2    \\
CPP-Nex  & $a$                           & 1.5      & 2.5        & 3    \\
         & $b$                           & 3        & 2          & 2    \\
         & $\omega^\star$                & 0.8      & 0.8        & 0.9  \\
EXNEX    & $\phi$                        & 0.125    & 0.125      & 2    \\
         & $w$                           & 0.3      & 0.2        & 0.6  \\
Fujikawa & $\varepsilon$                 & 0.5      & 2          & 3    \\
         & $\tau$                        & 0.4      & 0          & 0.2  \\
JSD-Global  & $\varepsilon$                 & 0.5      & 1          & 2    \\
         & $\varepsilon^\star$           & 2        & 2.5        & 1    \\
         & $\tau$                        & 0        & 0.3        & 0.5  \\ 
         \bottomrule
\end{tabular}
\end{table}

\clearpage
\newpage

\hypertarget{posterior-means}{%
\section{Posterior Means}\label{posterior-means}}

Mean posterior means are shown in Table \ref{tab:postmean}. Naturally, the methods that tend to borrow more information have slightly more bias in scenarios with different response rates. No unexpected differences are seen.

\begin{longtable}[]{cccccc}
\caption{Mean posterior means under all scenarios using the optimal tuning parameter values for each method}
\label{tab:postmean} \\
\toprule
Scenario & Method & Basket 1 & Basket 2 & Basket 3 & Basket 4 \\ \midrule
Global Null 
& BHM      & 0.150 & 0.149 & 0.149 & 0.149 \\ 
& BMA      & 0.155 & 0.154 & 0.155 & 0.155 \\ 
& CPP      & 0.161 & 0.161 & 0.161 & 0.161 \\ 
& CPP-Global  & 0.162 & 0.162 & 0.162 & 0.162 \\ 
& CPP-Nex  & 0.162 & 0.162 & 0.162 & 0.162 \\ 
& EXNEX    & 0.149 & 0.149 & 0.149 & 0.149 \\ 
& Fujikawa & 0.182 & 0.182 & 0.182 & 0.182 \\ 
& JSD-Global  & 0.163 & 0.162 & 0.163 & 0.163 \\ 
& MML      & 0.160 & 0.158 & 0.159 & 0.159 \\ 
\midrule
Global Alt 
& BHM      & 0.399 & 0.400 & 0.400 & 0.401 \\ 
& BMA      & 0.401 & 0.402 & 0.402 & 0.402 \\ 
& CPP      & 0.403 & 0.403 & 0.403 & 0.403 \\ 
& CPP-Global  & 0.404 & 0.404 & 0.404 & 0.404 \\ 
& CPP-Nex  & 0.404 & 0.404 & 0.404 & 0.404 \\ 
& EXNEX    & 0.399 & 0.400 & 0.400 & 0.401 \\ 
& Fujikawa & 0.409 & 0.409 & 0.409 & 0.409 \\ 
& JSD-Global  & 0.403 & 0.404 & 0.404 & 0.405 \\ 
& MML      & 0.402 & 0.403 & 0.403 & 0.404 \\ 
\midrule
One in the Middle 
& BHM      & 0.401 & 0.400 & 0.356 & 0.447 \\ 
& BMA      & 0.403 & 0.402 & 0.369 & 0.436 \\ 
& CPP      & 0.403 & 0.403 & 0.358 & 0.450 \\ 
& CPP-Global  & 0.404 & 0.404 & 0.359 & 0.449 \\ 
& CPP-Nex  & 0.404 & 0.404 & 0.359 & 0.449 \\ 
& EXNEX    & 0.401 & 0.401 & 0.365 & 0.437 \\ 
& Fujikawa & 0.409 & 0.409 & 0.362 & 0.456 \\ 
& JSD-Global  & 0.408 & 0.403 & 0.357 & 0.454 \\ 
& MML      & 0.404 & 0.403 & 0.335 & 0.474 \\ 
\midrule
Linear 
& BHM      & 0.223 & 0.271 & 0.325 & 0.381 \\ 
& BMA      & 0.232 & 0.278 & 0.331 & 0.375 \\ 
& CPP      & 0.234 & 0.280 & 0.332 & 0.384 \\ 
& CPP-Global  & 0.238 & 0.282 & 0.332 & 0.382 \\ 
& CPP-Nex  & 0.238 & 0.282 & 0.332 & 0.382 \\ 
& EXNEX    & 0.232 & 0.280 & 0.321 & 0.366 \\ 
& Fujikawa & 0.231 & 0.291 & 0.347 & 0.403 \\ 
& JSD-Global  & 0.233 & 0.280 & 0.341 & 0.403 \\ 
& MML      & 0.196 & 0.263 & 0.343 & 0.421 \\ 
\midrule
Good Nugget
& BHM      & 0.178 & 0.178 & 0.179 & 0.313 \\ 
& BMA      & 0.184 & 0.184 & 0.184 & 0.317 \\ 
& CPP      & 0.185 & 0.185 & 0.185 & 0.315 \\ 
& CPP-Global  & 0.189 & 0.189 & 0.189 & 0.311 \\ 
& CPP-Nex  & 0.189 & 0.189 & 0.189 & 0.311 \\ 
& EXNEX    & 0.183 & 0.182 & 0.183 & 0.300 \\ 
& Fujikawa & 0.198 & 0.198 & 0.198 & 0.347 \\ 
& JSD-Global  & 0.200 & 0.190 & 0.189 & 0.341 \\ 
& MML      & 0.170 & 0.170 & 0.171 & 0.373 \\ 
\midrule
Bad Nugget 
& BHM      & 0.242 & 0.370 & 0.372 & 0.371 \\ 
& BMA      & 0.248 & 0.373 & 0.374 & 0.373 \\ 
& CPP      & 0.256 & 0.379 & 0.379 & 0.379 \\ 
& CPP-Global  & 0.259 & 0.377 & 0.377 & 0.377 \\ 
& CPP-Nex  & 0.259 & 0.377 & 0.377 & 0.377 \\ 
& EXNEX    & 0.249 & 0.368 & 0.369 & 0.368 \\ 
& Fujikawa & 0.242 & 0.392 & 0.392 & 0.392 \\ 
& JSD-Global  & 0.246 & 0.386 & 0.387 & 0.387 \\ 
& MML      & 0.203 & 0.388 & 0.390 & 0.389 \\ 
\midrule
Half
& BHM      & 0.205 & 0.205 & 0.345 & 0.345 \\ 
& BMA      & 0.208 & 0.207 & 0.352 & 0.351 \\ 
& CPP      & 0.215 & 0.215 & 0.350 & 0.350 \\ 
& CPP-Global  & 0.220 & 0.220 & 0.348 & 0.348 \\ 
& CPP-Nex  & 0.220 & 0.220 & 0.348 & 0.348 \\ 
& EXNEX    & 0.218 & 0.217 & 0.332 & 0.332 \\ 
& Fujikawa & 0.217 & 0.217 & 0.373 & 0.373 \\ 
& JSD-Global  & 0.212 & 0.211 & 0.369 & 0.368 \\ 
& MML      & 0.182 & 0.181 & 0.379 & 0.379 \\ 
\bottomrule
\end{longtable}

\end{document}